\documentclass[12pt]{amsart}
\usepackage[english]{babel}
\usepackage{amsmath,amsthm}
\usepackage{amsfonts}
\usepackage{graphicx}

\theoremstyle{definition}

\theoremstyle{remark}

\numberwithin{equation}{section}
\newtheorem{algorithm}{Algorithm}
\usepackage[usenames]{color}         %for colors

\begin{document}

\title[Randomized MCMC and coupled MCMC]{Coupled MCMC with a randomized acceptance probability}%
\author{Geoff K. Nicholls$^\dagger$, Colin Fox$^\ast$ and Alexis Muir Watt$^\dagger$}%
\address{
$\dagger$: Department of Statistics\\
1 South Parks Road\\
Oxford, OX1 3TG\\
UK \\
$\ast$: Department of Physics\\
University of Otago\\
NZ
}%
%\date{}%
% ----------------------------------------------------------------
\begin{abstract}
We consider Metropolis Hastings MCMC with target $\pi(\theta)$ in cases where the log of the ratio of target
distributions $D=\log(\pi(\theta')/\pi(\theta))$ is replaced by an estimator $\hat D(W)$. The estimator
is based on $m$ samples $W=(W_1,W_2,...,W_m)$
from an independent online Monte Carlo simulation.
Under some conditions on the distribution of $\hat D(W)$ the process resembles Metropolis Hastings MCMC
with a randomized transition kernel. When this is the case there is a correction to the estimated acceptance
probability which ensures that the target distribution remains the equilibrium distribution.
The simplest versions of the \emph{penalty method}
of Ceperley and Dewing 1999~\cite{Ceperley1999}, the \emph{universal algorithm} of Ball et al. 2003 \cite{Ball2003}
and the \emph{single variable exchange algorithm} of Murray et al. 2006 \cite{Murray2006} are special cases.
In many applications of interest the correction terms cannot be computed. We consider approximate versions
of the algorithms.
We show that on average $O(m)$ of the samples realized by a simulation approximating a randomized
chain of length $n$ are exactly the same as those of a coupled (exact) randomized chain.
Approximation biases Monte Carlo estimates with terms $O(1/m)$ or smaller. This should be compared to
the Monte Carlo error which is $O(1/\sqrt{n})$.

% We give examples of approximate chains in which this bias decreases
%like $(1/m)$ and $1/m^{3/2}$ and show that the realize.
%We couple two approximate chains to the Metropolis algorithm and randomized Metropolis MCMC.
%The approximate chains realize just the same samples as the exact chains up to a random separation time $T$.
%The expected separation times with the Metropolis algorithm are both $O(\sqrt{m})$. However,
%the expected separation times with randomized MCMC are $O(m)$ and $O(m^{3/2})$.

\end{abstract}
\maketitle
% ----------------------------------------------------------------
\section{Introduction}

Monte Carlo simulation offers a direct route
to statistical inference for many otherwise awkward fitting problems.
The class of problems which may be treated using
Markov chain Monte Carlo (MCMC) and Sequential Monte Carlo has grown
a great deal since the core algorithms were proposed \cite{metropolis53,gordon93}.
One of the most important recent advances \cite{lin00, beaumont03, andrieu09, andrieu10} has given us
pseudo-marginal MCMC algorithms which are useful for some doubly intractable distributions.
Such distributions are hard to simulate as we cannot readily compute the ratio
of densities at two values of the target variable. Algorithms with a pseudo-marginal target distribution
put an estimate for the target distribution in the Monte Carlo state along with the target variable. In this paper we
give a class of MCMC algorithms characterized by a pseudo-marginal transition probability kernel.
In these algorithms the Monte Carlo state is just
the target variable. The class as a whole seems to be unknown
in statistical inference, though specific examples have appeared in various contexts in the
physics literature.

The MCMC algorithms we describe are,
for the most part, exact only in the case that the ratio estimator has a known
log-normal distribution. By `exact' we mean ergodic, with equilibrium distribution
equal the desired target, not `perfect' in the sense of Propp and Wilson 1996 \cite{propp96} and 1998 \cite{propp98}.
The restriction to log-normal estimators corresponds to the case where we have a normal estimator for
the log-likelihood. We know of no real problem which presents this
feature. However, a ratio estimator may be approximately log-normal when there is an appropriate CLT.
We show that an inexact MCMC algorithm may be adjusted to yield,
with high probability, just the same samples as a suitably coupled exact algorithm, in any
simulation of fixed length. At any fixed precision of the overall Monte Carlo estimate
the approximation error may be zero.

The paper has five sections. In Section~\ref{sec:rmcmc} we give a class of Metropolis Hastings
MCMC algorithms with a pseudo-marginal transition kernel. This is randomized MCMC. In Section~\ref{sec:examples} we show that
three existing MCMC algorithms for doubly intractable problems are special cases
of the algorithm described in Section~\ref{sec:rmcmc}. In Section~\ref{sec:cs} we give
a coupling-separation algorithm motivating the use of approximate MCMC. In Section~\ref{sec:application}
we give two very simple examples, chosen so that the coupling separation algorithm can
be implemented exactly. We conclude with a brief discussion of the results in this paper.

\section{Randomized Metropolis Hastings algorithms}
\label{sec:rmcmc}

%The general idea is to take a standard MCMC algorithm (which would not be useful for a
%doubly intractable target) and introduce an additional independent source of
%random variation within the Hastings ratio . If this

\subsection{Standard MCMC}
We begin with the standard MCMC algorithm of Hastings 1970 \cite{hastings70} and Metropolis et al. 1955 \cite{metropolis53}.
We call this the s-algorithm (standard algorithm). Our notation follows Tierney \cite{tierney94}. %There are many algorithms in this paper.

Let $\theta\sim \pi(d\theta)$ be a target variable with state space $E$, having a distribution $\pi(d\theta)=\pi(\theta)\mu(d\theta)$
which has a density $\pi(\theta)$ with respect to a measure $\mu(d\theta)$ defined for sets in a sigma-algebra $\mathcal E$ of subsets
of $E$. Let $Q(\theta,d\theta')=q(\theta,\theta')\mu(d\theta')$ be a Hastings proposal distribution
with density $q$ with respect to $\mu$, satisfying $q(\theta,\theta')>0\Leftrightarrow q(\theta',\theta)>0$. Let
\[
h(\theta,\theta')=\frac{\pi(\theta')q(\theta',\theta)}{\pi(\theta)q(\theta,\theta')}
\]
so that
\[
\alpha(\theta,\theta')=\min\left\{1,h(\theta,\theta')\right\}
\]
is the standard Metropolis Hastings acceptance probability. Let
\begin{equation}\label{eq:tp}
p(\theta,\theta')=\left\{\begin{array}{cc}
                           q(\theta,\theta')\alpha(\theta,\theta') & \theta\ne \theta' \\
                           0 & \theta=\theta'
                         \end{array}\right.
\end{equation}
be the zeroed transition probability density
in the s-algorithm. Let
\[
r(\theta)=1-\int_E p(\theta,\theta')\mu(d\theta')
\]
give the probability for a rejection,
and let $\mathbb{I}_{\theta=\theta'}$ be the indicator function for the event $\theta=\theta'$.
The transition probability distribution for s-MCMC is
\[
P(\theta,d\theta')=p(\theta,\theta')\mu(d\theta')+r(\theta)\mathbb{I}_{\theta=\theta'}.
\]

\subsection{Randomized MCMC}
\label{sec:ramcmc}
We now give algorithms with a randomized acceptance probability.
We call these algorithms r-algorithms.
%Consider a family of
%proposal distributions $Q(\theta,d\theta'; x)$ indexed by $x\in W$. In the setting of
%interest, $x$ is a real scalar, and $W$ is $\Re$. The proposal distributions
%are paired, so that a function $f(x)$, $f:W\rightarrow W$ gives the index of the proposal
%distribution $Q(\theta,d\theta'; f(x))$ reversing the one at $x$.
%We need $f$ to be an involution of the reals, so that $f(f(x))=x$.

Let $X$ be a real scalar random variable with probability density $\xi(x;\theta,\theta')$.
We assume the support $W$ of $\xi$ is independent of $\theta$ and $\theta'$. Let $f:W\rightarrow W$ be an
\emph{involution}, i.e. $f$ satisfies $f(f(x))=x$. The involution $f$ can be thought of as pairing points $(x,f(x))$ in $W$. We assume that $f$ has a derivative at $\xi$-a.e. $x\in W$.
 Examples of suitable involutions are the trivial involution $f(x)=x$, and
the family of functions
\[
f(x)=\frac{ax+b}{cx-a}
\]
with $a^2+bc\ne 0$.

%Besag et al 1995~\cite{besag95} show that continuously indexed proposal distributions, each
%satisfying detailed balance on its own, can be used in MCMC.
%
%We indexed the proposals in pairs $(x,f(x))$
%satisfying detailed balance together. \cf Pretty sure this is not true cf\
%
%The possible dependence of the density of $X$ on $\theta'$, the candidate state, is
%useful below.

Let
\[
h_{\xi}(\theta,\theta';x)=h(\theta,\theta')\frac{\xi(f(x);\theta',\theta)}
                                             {\xi(x;\theta,\theta')}|f'(x)|
\]
so that
\begin{equation}\label{eq:axi}
\alpha_{\xi}(\theta,\theta';x)=\min\left\{1,h_{\xi}(\theta,\theta';x)\right\}
\end{equation}
gives an acceptance probability which is randomized by $X$.
%For a given proposal distribution $Q$,
% all the subsequent distributions defined above, which depend on the acceptance probability, are also $\xi$-dependent.

The r-algorithm is as follows.
%
%{\sf \noindent
%Randomized-acceptance MCMC algorithm\\
%Let $\Theta_t=\theta$. The value of $\Theta_{t+1}$ is determined in the following way.\\
%1. Simulate $\theta'\sim q(\theta,\cdot)$ and $x\sim \xi(\cdot;\theta,\theta')$.\\
%2. With probability $\alpha_\xi(\theta,\theta';x)$ set $\Theta_{t+1}=\theta'$, otherwise
%set $\Theta_{t+1}=\theta$}.
%
\begin{algorithm}[r-MCMC]
\label{alg:rmcmc}
At state $\Theta_t=\theta$, simulate $\Theta_{t+1}$ as follows:
  \begin{enumerate}
    \item Simulate $\theta'\sim q(\theta,\cdot)$ and $x_t\sim \xi(\cdot;\theta,\theta')$.
    \item With probability $\alpha_\xi(\theta,\theta';x_t)$ set $\Theta_{t+1}=\theta'$ and otherwise
set $\Theta_{t+1}=\theta$.
\end{enumerate}
\end{algorithm}

We now show that the r-algorithm simulates a transition kernel that is in detailed balance with $\pi$,
so that
$\left\{\Theta_t\right\}_{t=0,1,2,...}$ is a Markov chain targeting $\pi$.
The acceptance probability in the r-algorithm is
\begin{equation}\label{eq:apxi}
\alpha_\xi(\theta,\theta')=\int_W \xi(x;\theta,\theta') \alpha_{\xi}(\theta,\theta';x) dx
\end{equation}
and we wish to establish detailed balance, i.e.,
\begin{equation}\label{eq:db}
\pi(\theta)q(\theta,\theta')\alpha_\xi(\theta,\theta')=\pi(\theta')q(\theta',\theta)\alpha_\xi(\theta',\theta).
\end{equation}
Multiplying both sides of Eqn~\ref{eq:axi} by $\pi(\theta)q(\theta,\theta')\xi(x;\theta,\theta')$ shows that
\begin{eqnarray} \label{eq:vdb1}
  &&\pi(\theta)q(\theta,\theta')\xi(x;\theta,\theta')\alpha_{\xi}(\theta,\theta';x) = \\
  && \min\left\{\pi(\theta)q(\theta,\theta')\xi(x;\theta,\theta'),\pi(\theta')q(\theta',\theta)\xi(f(x);\theta',\theta)|f'(x)|\right\}.
  \nonumber
\end{eqnarray}
Similarly,
\begin{eqnarray} \label{eq:vdb2}
  &&\pi(\theta')q(\theta',\theta)\xi(y;\theta',\theta)\alpha_{\xi}(\theta',\theta;y) = \\
  && \min\left\{\pi(\theta')q(\theta',\theta)\xi(y;\theta',\theta),\pi(\theta)q(\theta,\theta')\xi(f(y);\theta,\theta')|f'(y)|\right\}.
  \nonumber
\end{eqnarray}
In Eqn~\ref{eq:vdb2} set $y=f(x)$ so $x=f(y)$ and $f'(y)=1/f'(x)$ by the inverse function theorem. It follows that
\begin{eqnarray} \label{eq:vdb3}
  &&\pi(\theta')q(\theta',\theta)\xi(f(x);\theta',\theta)\alpha_{\xi}(\theta',\theta;f(x))|f'(x)| = \\
  && \min\left\{\pi(\theta')q(\theta',\theta)\xi(f(x);\theta',\theta)|f'(x)|,\pi(\theta)q(\theta,\theta')\xi(x;\theta,\theta')\right\}
  \nonumber
\end{eqnarray}
and this equation has RHS equal to the RHS of Eqn~\ref{eq:vdb1}. It follows that the LHS of Eqns~\ref{eq:vdb1} and \ref{eq:vdb3} are equal, i.e.,
\begin{eqnarray}\label{eq:vdb}
&&\pi(\theta)q(\theta,\theta')\alpha_{\xi}(\theta,\theta';x)\xi(x;\theta,\theta') = \\
                    &&\hspace{1in}\pi(\theta')q(\theta',\theta)\alpha_{\xi}(\theta',\theta;f(x))\xi(f(x);\theta',\theta)|f'(x)|.  \nonumber
\end{eqnarray}
This is `very detailed balance'. Integrating Eqn~\ref{eq:vdb} over all $x$ in $W$ gives detailed balance on average, Eqn~\ref{eq:db}.
%It is interesting to note that each side of the  equivalent relation to Eqn~\ref{eq:vdb} for the standard Metropolis-Hastings iteration is symmetric in $\theta$ and $\theta'$ (see e.g.~\cite{liu05} p. 113) which establishes detailed balance directly. In general each side of  Eqn~\ref{eq:vdb} is not symmetric in $\theta$ and $\theta'$, in which case the kernel simulated by each iteration of the r-algorithm is not in detailed balance with $\pi$. However, as we have shown, the r-algorithm satisfies detailed balance \emph{on average}, and hence targets $\pi$. Thus
%The r-algorithm differs from the scheme of
Besag et al 1995 \cite{besag95} show that continuously indexed proposal distributions, each satisfying detailed balance on its own, can be used in MCMC. In our setup the transition kernels satisfy detailed balance in pairs, the kernel at $x$ with the kernel at $f(x)$.

In the next section we mention algorithms with more than one randomization. We claim without proof that
the properties of r-MCMC established in this and the next sub-section hold under the generalization from scalar to multivariate $X$.
That is, if $X=(X_1,...,X_K)$ has multivariate density $\xi(x;\theta,\theta')$ with support $W$ in $\Re^K$,
and $f$ is an involution of $W$ having Jacobian $f'(x)$ with determinant $|f'(x)|$, then the r-algorithm in Alg~\ref{alg:rmcmc}
targets $\pi$.

\subsection{Properties of r-MCMC}

In this section and Appendices A and B we show that r-MCMC is less statistically efficient than the s-MCMC from which it is derived,
and give a sufficient condition for the r-chain to inherit $\pi$-irreducibility and minorization from the corresponding s-chain.
These results can be used to establish ergodicity in particular cases.

First, r-MCMC is less statistically efficient than the s-MCMC.
We show (in Appendix A) that, for all $\theta$, $\theta'\in E$,
\begin{equation}\label{eq:apub}
    \alpha_{\xi}(\theta,\theta')\le \alpha(\theta,\theta').
\end{equation}
Since the two chains have the same proposal distribution $Q(\theta,d\theta')$, this puts the
r-chain below the s-chain in the ordering of Peskun (1973) \cite{peskun73}, and so r-chain estimators have greater asymptotic
variance than corresponding s-chain estimators.

We next give a sufficient condition for r-chain $\pi$-irreducibility and minorization.
Let $p_\xi(\theta,\theta')$ be given by Eqn~\ref{eq:tp}, with $\alpha(\theta,\theta')$ replaced by $\alpha_\xi(\theta,\theta')$
(see Appendix B for details).
Let $r_\xi(\theta)=1-\int_E p_\xi(\theta,\theta')\mu(d\theta')$ and
\[
P_\xi(\theta,d\theta')=p_\xi(\theta,\theta')\mu(d\theta')+r_\xi(\theta)\mathbb{I}_{\theta=\theta'}.
\]
Suppose there exists a constant $\epsilon>0$ such that
\begin{equation}\label{eq:epsbound}
\Pr\left(\frac{\xi(f(X);\theta',\theta)}{\xi(X;\theta,\theta')}|f'(X)|\ge 1\right)\ge\epsilon,
\end{equation}
for $X\sim \xi(\cdot;\theta,\theta')$, independent of $\theta$ and $\theta'$ in $E$.
If $\xi$ does not depend on $\theta,\theta'$, as in the penalty method
example~\cite{Ceperley1999} below, condition Eqn~\ref{eq:epsbound} is automatically satisfied (see Appendix B).

We show (in Appendix B) that if Eqn~\ref{eq:epsbound} holds then
\begin{equation}\label{eq:tplb}
    P_\xi(\theta,C)\ge \epsilon P(\theta,C).
\end{equation}
Because the s-chain is $\pi$-irreducible by assumption, for each $C\in \mathcal E$ such that
$\pi(C)>0$ and for $\pi$-a.e. $\theta\in E$ there exists $n=n(\theta,A)$ such that $P^{(n)}(\theta,C)>0$. However,
if Eqn~\ref{eq:tplb} holds then $P^{(n)}_\xi(\theta,C)\ge \epsilon^n P^{(n)}(\theta,C)$,
so the r-chain is $\pi$-irreducible also.

Suppose the s-chain satisfies a minorization
condition $M(m,\beta,C,\nu)$. This means that there exist $m\ge 1$, $\beta>0$, a set $C\in \mathcal E$ and
a probability measure $\nu$ on $\mathcal E$ such that $\nu(C)>0$ and $P^{(m)}(x,B)\ge \beta\nu(B)$
for all $x\in C$ and all $B\in \mathcal E$. It follows from Eqn~\ref{eq:tplb} that the r-chain satisfies a minorization
condition $M(m,\epsilon^m\beta,C,\nu)$. If the s-chain is uniformly ergodic, then so
is the r-chain.

%We have described a family of randomized Metropolis Hastings algorithms and shown that they
%inherit certain properties of the Metropolis Hastings algorithm from which they are derived.

\subsection{Example}
\label{sec:simpleexample}
We now give a simple example illustrating the r-algorithm. Suppose we have an s-algorithm
targeting $\pi(\theta)$ with symmetric proposal $q(\theta,\theta')=q(\theta',\theta)$ and acceptance probability
\[
\alpha(\theta,\theta')=\min\left\{1,\frac{\pi(\theta')}{\pi(\theta)}\right\}.
\]
We can randomize this acceptance with a normal density $\xi(x;\theta,\theta')=N(x;a,b)$ having mean $a=\log(\pi(\theta')/\pi(\theta))$ and
variance $b=1$, and use the identity involution $f(x)=x$, to get the following r-algorithm:
\begin{algorithm}[Example]
\label{alg:toy}
At state $\Theta_t=\theta$, simulate $\Theta_{t+1}$ as follows:
  \begin{enumerate}
    \item Simulate $\theta'\sim q(\theta,\cdot)$ and $z_t\sim N(0,1)$. Set $$x_t=\log(\pi(\theta')/\pi(\theta))+z_t.$$
    \item With probability
\[
\alpha_\xi(\theta,\theta';x_t) = \min\left\{1,\left(\frac{\pi(\theta')}{\pi(\theta)}\right)^{1-2x_t}\right\}
\]
set $\Theta_{t+1}=\theta'$, otherwise set $\Theta_{t+1}=\theta$.
\end{enumerate}
\end{algorithm}
%
%{\sf \noindent
%Randomized MCMC algorithm: toy problem\\
%Let $\Theta_t=\theta$. The value of $\Theta_{t+1}$ is determined in the following way.\\
%1. Simulate $\theta'\sim q(\theta,\cdot)$ and $Z\sim N(0,1)$. Set $X=\log(\pi(\theta')/\pi(\theta))+Z$.\\
%2. With probability
%\[
%\alpha_\xi(\theta,\theta';x) = \min\left\{1,\left(\frac{\pi(\theta')}{\pi(\theta)}\right)^{1-2X}\right\}
%\]
%set $\Theta_{t+1}=\theta'$ and otherwise set $\Theta_{t+1}=\theta$.}\\[0.1in]
The example algorithm satisfies detailed balance with respect to $\pi$.
The correction term can be interpreted as
kind of random tempering, since the random power flattens the target distribution when $0\le 1-2x_t<1$. However this `tempering' does not decrease the integrated autocorrelation time of the Markov chain. As shown above, the r-chain is dominated by the s-chain in the Peskun ordering, and hence the integrated autocorrelation time is not decreased.

\section{Known randomized Metropolis Hastings algorithms}
\label{sec:examples}

We now describe some existing MCMC algorithms for doubly intractable distributions
and show that they are r-algorithms. From this point on, we assume for ease of exposition that the proposal density is symmetric, $q(\theta,\theta')=q(\theta',\theta)$, since the Hastings extensions are straightforward.
Suppose $$D(\theta,\theta')=\log(\pi(\theta')/\pi(\theta))$$ is an intractable function of $(\theta,\theta')$.
The acceptance probability in the s-algorithm for $\pi(\theta)$ is $$\alpha(\theta,\theta')=\min(1,\exp(D(\theta,\theta'))).$$
Let $\hat D_{\theta,\theta'}=\hat D_{\theta,\theta'}(W)$
be an estimator for $D(\theta,\theta')$ computed from a collection of $m$
random variables $W=(W_1,W_2,...,W_m)$. Estimator
$\hat D_{\theta,\theta'}$ has cdf $G_m(\cdot;\theta,\theta')$
and density $g_m(\cdot;\theta,\theta')$.
Let
$$\sigma^2 =\lim_{m\rightarrow \infty}\mbox{var}(\sqrt{m}\hat D_{\theta,\theta'}(W)),$$
so that the estimaor variance is asymptotically $\sigma^2/m$.
For example, if the $W_i$ are iid, and $\hat D_{\theta,\theta'}(W)=\overline W$,
then $\sigma^2=\mbox{var}(W_1)$.

We begin with a `naive' incorrect algorithm which is not itself an r-algorithm.
We refer to an algorithm as \emph{naive} when an estimate is plugged into the Metropolis Hastings acceptance probability, without correction.
The `MCWM' algorithm in Andrieu and Roberts 2009~\cite{andrieu09} is an algorithm in this class.
Although inexact, the algorithm may be useful. Beaumont 2003 \cite{beaumont03}
has shown that the approximation bias may be small, and the resulting (inexact) chain seems to have better mixing properties
than the corresponding (exact) pseudo-marginal MCMC.
\begin{algorithm}[Naive algorithm]
\label{alg:naive}
At state $\Theta_t=\theta$, simulate $\Theta_{t+1}$ as follows:
  \begin{enumerate}
    \item Simulate $\theta'\sim q(\theta,\cdot)$ and an estimate $x_t \sim g_m(\cdot;\theta,\theta')$.
    \item With probability
\[
\alpha_\text{N}(\theta,\theta')=\min\left\{1,\mathrm{e}^{x_t}\right\}
\]
set $\Theta_{t+1}=\theta'$, otherwise set $\Theta_{t+1}=\theta$.
\end{enumerate}
\end{algorithm}
The naive algorithm does not in general target $\pi$, and may not even have an equilibrium distribution.

%{\sf \noindent
%Naive algorithm\\
%Let $\Theta_t=\theta$. Simulate $\Theta_{t+1}$ as follows.\\
%1. Simulate $\theta'\sim q(\theta,\cdot)$ and an estimate $x \sim G_m(\cdot;\theta,\theta')$.\\
%2. With probability
%\[
%\alpha_N(\theta,\theta')=\min\left\{1,e^{x}\right\}
%\]
%set $\Theta_{t+1}=\theta'$, otherwise set $\Theta_{t+1}=\theta$.}\\[0.1in]

The penalty method of Ceperley and Dewing 1999~\cite{Ceperley1999}
corrects the acceptance probability in the naive algorithm. A variant of this algorithm is
used by these and other authors to simulate Gibbs distributions of interest in physical chemistry.
%The method corrects for the error which arises when an estimate for the difference in Gibbs potentials
%is plugged into the Metropolis algorithm in place of the exact difference.
%In order to make possible statistical applications clear, we recast the problem in a Bayesian setting,
%with data $y$, prior $p(\theta)$ and log-likelihood $\ell(\theta;y)$. Temporarily suppressing the data-dependence
%of $\ell$, the target
%posterior distribution is \[\pi(\theta|y))\propto p(\theta)\exp(\ell(\theta)).\]
%Let $D(theta,\theta')=\ell(\theta')-\ell(\theta)$.
Following Ceperley \& Dewing 1999~\cite{Ceperley1999}), we
suppose a normal estimator $\hat D_{\theta,\theta'}\sim N(D(\theta,\theta'),\sigma^2(\theta,\theta')/m)$ is available.
We assume $\sigma(\theta,\theta')=\sigma(\theta',\theta)$ and
write $\sigma=\sigma(\theta,\theta')$ below.
Ceperley \& Dewing 1999~\cite{Ceperley1999} show that the following algorithm, which they call the `penalty method',
targets $\pi(\theta)$ exactly.
\begin{algorithm}[Penalty method (Ceperley \& Dewing 1999~\cite{Ceperley1999})]
\label{alg:penalty}
At state $\Theta_t=\theta$, simulate $\Theta_{t+1}$ as follows:
  \begin{enumerate}
    \item Simulate $\theta'\sim q(\theta,\cdot)$ and an estimate $y_t \sim N(D(\theta,\theta'),\sigma^2/m)$.
    \item With probability
\[
\alpha_P(\theta,\theta')=\min\left\{1,e^{y_t-\sigma^2/2m}\right\}
\]
set $\Theta_{t+1}=\theta'$, otherwise set $\Theta_{t+1}=\theta$.
\end{enumerate}
\end{algorithm}

%{\sf \noindent
%Penalty method (Ceperley \& Dewing 1999~\cite{Ceperley1999})\\
%Let $\Theta_t=\theta$. Simulate $\Theta_{t+1}$ as follows.\\
%1. Simulate $\theta'\sim q(\theta,\cdot)$ and an estimate $y \sim N(D(\theta,\theta'),\sigma^2)$.\\
%2. With probability
%\[
%\alpha_P(\theta,\theta')=\min\left\{1,e^{y-\sigma^2/2}\right\}
%\]
%set $\Theta_{t+1}=\theta'$, otherwise set $\Theta_{t+1}=\theta$.}\\[0.1in]

Ceperley and Dewing 1999~\cite{Ceperley1999} show that detailed balance is satisfied by carrying out the integrals
over $y_t$ needed to verify Eqn~\ref{eq:db}.
We now show that the penalty method is an r-algorithm.
Let $X_t$ have a normal density, $\xi(x;\theta,\theta')=N(x;0,\sigma^2m)$ and take for
$f$ the involution $f(x)=\sigma^2-x$. It follows that
\[
\frac{\xi(f(x_t);\theta',\theta)}{\xi(x_t;\theta',\theta)}|f'(x_t)|=\mathrm{e}^{x_t-\sigma^2/2m}
\]
and hence the acceptance probability in the r-algorithm at step $t$ is
\[
\alpha_\xi(\theta,\theta';x_t)\min\left\{1,\mathrm{e}^{D(\theta,\theta')+x_t-\sigma^2/2m}\right\}.
\]
Since $D(\theta,\theta')+X_t$ (in the r-algorithm) and $Y_t$ (in the penalty method in Alg~\ref{alg:penalty}) have the same distribution,
 the algorithms are equivalent and the results in Sec~\ref{sec:ramcmc} prove that the penalty method targets $\pi$.
 %The penalty method provides an efficient implementation of the r-algorithm since only the estimator $\hat D_{\theta,\theta'}$ is needed, rather than both $D(\theta,\theta')$ and $x$ as stated in the r-algorithm in Alg~\ref{alg:rmcmc}.

 Ceperley and Dewing 1999~\cite{Ceperley1999} give other more general but approximate algorithms, the simplest of
 which is the Penalty Estimate method. We analyze this alongside the naive algorithm in Section~\ref{sec:cs}.
 Suppose we have a unbiased normal estimator for $D$, but do not
know its variance, and so try replacing $\sigma^2$ with the sample variance $s^2$.
\begin{algorithm}[Penalty Estimate method (Ceperley \& Dewing 1999~\cite{Ceperley1999})]
\label{alg:penalty}
At state $\Theta_t=\theta$, simulate $\Theta_{t+1}$ as follows:
  \begin{enumerate}
    \item Simulate $\theta'\sim q(\theta,\cdot)$, an estimate $y_t \sim N(D(\theta,\theta'),\sigma^2/m)$
    and an independent variance estimate $s^2$ with $$(m-1)s^2_t/\sigma^2\sim \chi^2(m-1).$$
    \item With probability
\[
\alpha_{\hat P}(\theta,\theta')=\min\left\{1,e^{y_t-s_t^2/2m}\right\}
\]
set $\Theta_{t+1}=\theta'$, otherwise set $\Theta_{t+1}=\theta$.
\end{enumerate}
\end{algorithm}
This algorithm is inexact. It is not an r-algorithm.

%MCMC algorithms using estimators may be idenified as r-algorithms by  choosing the randomization $\xi(x;\theta,\theta')$
%and involution $f(x)$ so that the randomization mimics the distribution of the estimator, that is
%\[
%\hat D(W)\sim D + \log\left(\frac{\xi(f(X);\theta',\theta)}
%                                             {\xi(X;\theta,\theta')}|f'(X)|\right).
%\]

Ball et al. 2003~\cite{Ball2003} give a `universal rule' which
applies for symmetric error distributions with compact support and does not require that the error variance $\sigma^2$ be known. In the case
where the error distribution is normal the algorithm can target only a very good approximation to $\pi$.
We can show that the universal rule for a normal error distribution is an r-algorithm, with two randomizations,
$\xi_1(u_1;\theta,\theta')=N(u_1;0,\sigma^2/m)$ and $\xi_2(u_2;\theta,\theta')=N(u_2;\sigma^2/m,\sigma^2/m)$ and the same involution,
$f(u_i)=\sigma^2/m-u_i, i=1,2$, for each. In this bivariate randomization the unknown variance appears in a factor multiplying
the whole acceptance probability. The acceptance probability may be simulated by a form of rejection. Promising though it is, we do not discuss
the universal algorithm here, as there is no exact algorithm for the normal case and we make no further use of the connection.

The final algorithm we discuss is the \emph{single variable exchange} algorithm of Murray \& MacKay 2006~\cite{Murray2006}.
We present this in its original Bayesian setting,
with data $d=(d_1,...,d_n)$, prior $p(\theta)$ and Likelihood $L(\theta,d)=c(\theta)\tilde L(\theta,d)$. The target is
now the posterior distribution, \[\pi(\theta|d))\propto p(\theta)L(\theta,d).\]
The likelihood has an intractable normalizing constant $c(\theta)$. Murray \& MacKay 2006~\cite{Murray2006}
arrange things so that this cancels in the Hastings ratio, developing an idea due to M{\o}ller et al. 2004~\cite{Moller2004}.
However, while M{\o}ller et al. 2004~\cite{Moller2004} augment the MCMC state with an auxiliary variable,
the auxiliary variable in the exchange algorithm is associated with a single update,
as in the r-algorithm.
\begin{algorithm}[Single Variable Exchange]
\label{alg:sve}
At state $\Theta_t=\theta$, simulate $\Theta_{t+1}$ as follows:
  \begin{enumerate}
    \item Simulate $\theta'\sim q(\theta,\cdot)$ and $x_t\sim L(\theta',\cdot)$.
    \item With probability
\[
\alpha_E(\theta,\theta')=\min\left\{1,\frac{p(\theta')L(\theta',d)}{p(\theta)L(\theta,d)}\frac{L(\theta,x_t)}{L(\theta',x_t)}\right\}
\]
set $\Theta_{t+1}=\theta'$ and otherwise set $\Theta_{t+1}=\theta$.
\end{enumerate}
\end{algorithm}
%{\sf \noindent
%Single Variable Exchange\\
%If $\Theta_t=\theta$ then $\Theta_{t+1}$ is determined in the following way.\\
%1. Simulate $\theta'\sim q(\theta,\cdot)$ and $x\sim L(\theta',\cdot)$.\\
%2. With probability
%\[
%\alpha_E(\theta,\theta')=\min\left\{1,\frac{p(\theta')L(\theta',y)}{p(\theta)L(\theta,y)}\frac{L(\theta,x)}{L(\theta',x)}\right\}
%\]
%set $\Theta_{t+1}=\theta'$ and otherwise set $\Theta_{t+1}=\theta$.}\\[0.1in]
Factors of $c(\theta)$ and $c(\theta')$ cancel. This is an r-algorithm with
$\xi(x;\theta,\theta')=L(\theta',x)$ and the identity involution $f(x)=x$.
This is the first useful instance we have given for r-MCMC (ie, excluding the example in Section~\ref{sec:simpleexample}) in which $\xi$ actually depends on $\theta'$.

How do these identifications of existing algorithms as r-algorithms help us? Both Ceperley \& Dewing 1999~\cite{Ceperley1999} and Ball et al. 2003~\cite{Ball2003} treat
detailed balance as an integral equation, integrated over the random variation introduced by the
estimator, $\hat D$. The acceptance probability is obtained as a solution of this integral
equation. In fact the 'very detailed' balance relation Eqn~\ref{eq:vdb} shows that the functions under the
integrals are equal. We expect that this will help with the development of new algorithms.

\section{Separation times and approximate-target MCMC}
\label{sec:cs}

In this section we give a coupling strategy which motivates the use of the naive algorithm
in some cases.
We present and analyze a coupling algorithm to show that, on average, the naive algorithm gives {\it exactly the same} MCMC samples as the exact penalty method, out to $O(m)$ steps of the naive chain, where $m$ is the sample size used in $D$-estimation in the naive chain.
The error in this algorithm is analyzed in Andrieu and Roberts 2009~\cite{andrieu09} from a
different perspective.

Recall that $\hat D_{\theta,\theta'}$ is an estimator for $D(\theta,\theta')$ with
cdf $G_m(\cdot;\theta,\theta')$.
We do not assume $\hat D_{\theta,\theta'}$ is unbiased or normal.
We do assume it satisfies a CLT, so that
\begin{equation}\label{eqn:cdfconverge}
G_m(x)=\Phi\left(\frac{x-D}{\sigma/\sqrt{m}}\right)+O(m^{-1/2}),
\end{equation}
with $\sigma^2$ the asymptotic variance of $\sqrt{m} \hat D_{\theta,\theta'}$.
For example, if $\hat D_{\theta,\theta'}$ is computed from a realization of a geometrically
ergodic Markov chain $W=\{W_i\}_{i=0}^\infty$ and
\begin{equation}\label{eq:geom}
\hat D_{\theta,\theta'}=\frac{1}{m}\sum_{i=1}^m W_i,
\end{equation}
then Eqn~\ref{eqn:cdfconverge} holds, subject to mild additional conditions
specified in Kontoyiannis and Meyn (2003) \cite{Kontoyiannis03}.
%The leading $O(m^{-1/2})$ terms
%are zero if the bias and skew are zero.
Results of this
kind may be used with the delta method to get asymptotically normal ratio estimators.
If Eqn~\ref{eqn:cdfconverge} holds for $\hat D_{\theta,\theta'}$, then it can be coupled to a normal estimator.
\begin{eqnarray}
\hat D_{\theta,\theta'}&=&D+\frac{\sigma}{\sqrt{m}}\frac{\hat D-D}{\sigma/\sqrt{m}} \nonumber \\
        &=&D+\frac{\sigma}{\sqrt{m}}\Phi^{-1}(G_m(\hat D)+O(m^{-1/2})) \nonumber \\
        &=&D+\frac{\sigma}{\sqrt{m}}\Phi^{-1}(G_m(\hat D))+O(1/m),\label{eq:coupling}
\end{eqnarray}
where $\Phi^{-1}(G_m(\hat D))$ is a standard normal random variable.

We now give the coupling algorithm. In our example, we couple the naive and penalty method chains.
Couplings of this kind may be applied to other pairs of algorithms.
The algorithm simulates the penalty method and also an indicator variable $B_t\in \{0,1\},t=1,2,...$
marking the times at which the naive chain separates from the penalty-method chain.
\begin{algorithm}[Coupling algorithm: penalty method and naive algorithm]
\label{alg:pmna}
At state $\Theta_t=\theta$, simulate $B_{t}$ and $\Theta_{t+1}$ as follows:
  \begin{enumerate}
    \item Simulate $\theta'\sim q(\theta,\cdot)$ and an estimate $x_t \sim g_m(\cdot;\theta,\theta')$,
and set $$y_t=D+\frac{\sigma}{\sqrt{m}}\Phi^{-1}(G_m(x_t)).$$
    \item Simulate $V_t\sim U(0,1)$. Let
       \[
          \alpha_\text{P}(\theta,\theta';y_t)=\min\left\{1,\mathrm{e}^{y_t-\sigma^2/2m}\right\}.
       \]
       If $V_t\le \alpha_\text{P}$ then set $\Theta_{t+1}=\theta'$, otherwise set $\Theta_{t+1}=\theta$.
    \item Let
      \[
          \alpha_\text{N}(\theta,\theta';x_t)=\min\left\{1,\mathrm{e}^{x_t}\right\}.
       \]
      If $$\min(\alpha_\text{N},\alpha_\text{P})<V_t\le \max(\alpha_\text{N},\alpha_\text{P})$$
       then set $B_{t}=1$ and otherwise set $B_{t}=0$.
\end{enumerate}
\end{algorithm}
%{\sf \noindent
%Coupling algorithm, Penalty method and Naive algorithm\\
%Let $\Theta_t=\theta$. Simulate $B_{t}$ and $\Theta_{t+1}$ as follows.\\
%1. Simulate $\theta'\sim q(\theta,\cdot)$ and an estimate $x_t \sim G_m(\cdot;\theta,\theta')$,
%and set $$y_t=D+\frac{\sigma}{\sqrt{m}}\Phi^{-1}(G_m(x)).$$
%2. Simulate $V_t\sim U(0,1)$. Let
%\[
%\alpha_P(\theta,\theta';y_t)=\min\left\{1,e^{y_t-\sigma^2/m}\right\}.
%\]
%If $V_t\le \alpha_P$ then set $\Theta_{t+1}=\theta'$ and otherwise set $\Theta_{t+1}=\theta$.\\
%3. Let
%\[
%\alpha_N(\theta,\theta';x_t)=\min\left\{1,e^{x_t}\right\}.
%\]
%If $$\min(\alpha_N,\alpha_P)<V_t\le \max(\alpha_N,\alpha_P)$$
%then set $B_{t}=1$ and otherwise set $B_{t}=0$.
%}\\[0.1in]

The $\Theta_t$ chain targets $\pi$ exactly, as it is a penalty method chain.
The two chains (naive and penalty method) separate when $V_t$ falls between the two acceptance
probabilities, since then the chains make different accept/reject decisions.
If this first separation time is larger than the run length, then the
naive algorithm realizes the same samples as the penalty method, and the error from using an inexact
chain is undetectable at the overall MCMC precision. We emphasize that we cannot usually
implement the coupling algorithm as we cannot in general compute $\Phi^{-1}(G_m(\hat D_{\theta,\theta'}))$.

%The two chains have the same proposal distribution $Q(\t,d\t')$, start in the same
%initial state $\Theta^{(N)}_t=\Theta^{(P)}_t=\t_0$, and use the same $U(0,1)$ random
%number streams $U_t$ and $V_t$ to simulate candidate states and the accept-reject
%event.

%We couple the naive algorithm with acceptance probability $\alpha_N(\t,\t')$ to an
%exact penalty method algorithm with acceptance probability $\alpha_P(\t,\t')$.
%The two chains have the same proposal distribution $Q(\t,d\t')$, start in the same
%initial state $\Theta^{(N)}_t=\Theta^{(P)}_t=\t_0$, and use the same $U(0,1)$ random
%number streams $U_t$ and $V_t$ to simulate candidate states and the accept-reject
%event. The two chains separate at the first time $V_t$ falls between the two acceptance
%probabilities, when $\min(\alpha_N,\alpha_P)<V_t\le \max(\alpha_N,\alpha_P)$.

%Consider the sequence of times $T_i=\min\{t>T_{i-1}; B_t=1\}$ setting $T_0=0$.

We now give a lower bound on the mean time to separation, assuming that the chains start in equilibrium.
Let $T=\min\{t\ge 1; B_{t-1}=1\}$ be the first passage time to separation.
We assume $\Pr(T<\infty)=1$.
Next, condition the process on a separation at the first trial, $B_0=1$, set $T_0=0$ and let
$$T_i=\min\{t>T_{i-1}; B_t=1, i=1,2,3,...|B_0=1\}$$
be the sequence of separation return times.
The intervals $T_i-T_{i-1}$ are the random intervals between separation events.
The $B_t$ process is not in general a Markov chain, and the intervals are not iid.
However, conditional on $B_0=1$, the $B_t$ process is a stationary
discrete process, with interval-stationary separation return times.
Let $\rho=\mathbb{E}(T_1)$ be the mean separation return-time.
By Kac's Recurrence Theorem for a stationary discrete process, $\rho=1/\Pr(B_0=1)$. Let %Denote by $\mathbb{E}|\alpha_P-\alpha_N|$ the expectation
\[
\mathbb{E}|\alpha_P-\alpha_N|=\int_{E^2\times R} \! |\alpha_P(\theta,\theta';\delta)-\alpha_N(\theta,\theta';\delta)|\, g_m(\delta)d\delta Q(\theta,d\theta')\pi(d\theta).
\]
The mean separation return-time is
\begin{equation}\label{eq:prt}
    \rho=\frac{1}{\mathbb{E}|\alpha_P-\alpha_N|}.
\end{equation}
The separation time from the initialization $\Theta_0=\theta_0$ is
$\tau(\theta_0)=\mathbb{E}(T|\Theta_0=\theta_0)$. Let $\tau=\int_E \tau(\theta) \pi(d\theta)$ give the mean separation time starting in
equilibrium. This time is the expected first passage time to a separation event $B_t=1$ from an equilibrium start for $\Theta_0$. The return time of an interval-stationary process bounds the first passage time by $2\tau\ge\rho$, as the return time around a fixed time is length-biased,
and the fixed time is uniformly distributed in an interval between separation events.
It follows that the mean separation time $\tau$ grows at least linearly with increasing $m$, since $|\alpha_P-\alpha_N|=O(1/m)$
in Eqn~\ref{eq:prt}.
We have not bounded the separation time $\tau(\theta_0)$ from any particular start state.
However we assume $\tau(\theta_0)\simeq \tau$ as the the event $\alpha_P=\alpha_N=1$
occurs more frequently during convergence, and the chains cannot separate on these events.

If we couple the naive algorithm to the standard algorithm, s-MCMC, with acceptance probability
$\alpha$, we find $\mathbb{E}|\alpha-\alpha_N|=O(1/\sqrt{m})$.
The naive algorithm is therefore `closer' to the exact penalty method, in the `distance' $\mathbb{E}|\alpha_P-\alpha_\text{N}|$
than it is to the exact standard algorithm.
We improve the naive approximation to the target by improving the approximation $G_m(x)\simeq \Phi\left(\frac{x-D}{\sigma/\sqrt{m}}\right)$.

%%%%%%%%%%%%%%%%%%%%%%%%%%%%%%%%%%%%%%%%%%%%%%%%%%%%%%%%%%%%%%%%%%%%%%%%%%%

The Penalty Estimate method (Algorithm 5) has separation times of $O(m^{3/2})$ at the price of
stronger conditions on the distribution of the estimator $\hat D_{\theta,\theta'}$.
The estimator $x_t\sim N(D(\theta,\theta'),\sigma^2/m)$ in the Penalty Estimate method is exactly normal,
but we do not know $\sigma^2$,
and proceed as in Algorithm 5. The estimator is normal, so it can be coupled to
the (exact) Penalty Method using the trivial coupling $y_t=x_t$. Now since
$$\frac{(m-1)s^2/\sigma^2-(m-1)}{\sqrt{2(m-1)}}\stackrel{D}{\rightarrow} N(0,1),$$
the two acceptance probabilities $\alpha_{\hat P}$ and $\alpha_P$
differ by terms of $O(m^{3/2})$. We give an example with this behavior in Section~\ref{sec:application}.
%It may be possible to achieve separations times of $O(m^{3/2})$
%for $W_{t,i}, i=1,2,...,m$ a Geometric Markov chain, as Kontoyiannis and Meyn 2003 \cite{Kontoyiannis03} give the terms of $O(1/m)$ in %Eqn~\ref{eqn:cdfconverge} for
%this case, and these terms are again zero if the distribution of $m^{-1}\sum_i W_{t,i}$ has zero bias and skew.
%However, we need an estimator for $\sigma^2$ with suitable asymptotics, and this is not straightforward.
%%%%%%%%%%%%%%%%%%%%%%%%%%%%%%%%%%%%%%%%%%%%%%%%%%%%%%%%%%%%%%%%%%%%%%%%%%%%%

Finally we note that if the chains are independence samplers, the naive chain
and the penalty method chain may separate and then coalesce. This kind of coupling
is used in the `perfect simulation' algorithm of Murdoch et al. 1998~\cite{murdoch98}. The two chains
are offered the same candidate at each step, even after separation, and
they coalesce when they both accept. If the
time to coalesce is much smaller than the time to separate,
the naive algorithm may give a very good approximation indeed.

\section{Examples}
\label{sec:application}

In order to show the linear dependence of the separation time $\tau$ on $m$, we give a very simple example for which we can compute
$\Phi^{-1}(G_m(x))$. Let $\pi$ be an equal mixture of bivariate normals, with $\theta=(\theta_1,\theta_2)$ and
\[
\pi(\theta)=\frac{1}{2}\: MVN(\theta;\mu_1,\Sigma_1)+\frac{1}{2}MVN(\theta;\mu_2,\Sigma_2),
\]
$\mu_1=(3,3)^T$, $\mu_2=(6,6)^T$, $[\Sigma_a]_{i,i}=1$ for $a,i=1,2$, and $[\Sigma_1]_{1,2}=1/2$ and
$[\Sigma_2]_{1,2}=-0.5$.
If $D=\log(\pi(\theta')/\pi(\theta))$ and for $i=1,2,...,m$, $W_i\sim \mbox{Exp}(1)$ then we use
$\hat D=D-1+m/\sum_i W_i$ to estimate $D$. In this example $G_m(x)=F((x+1)/m; m,1)$ where
$F(\cdot; m,1)$ is the cdf of an inverse Gamma variable with shape $M$ and rate $1$, so that
$\mathbb{E}(\hat D)=D+1/(M-1)$ and $\mbox{var}(\hat D)=m^2/(m-1)^2(m-2)$. In this example,
$\hat D$ is a biased, non-normal estimator for $D$.

\begin{figure}
  \includegraphics[width=4in]{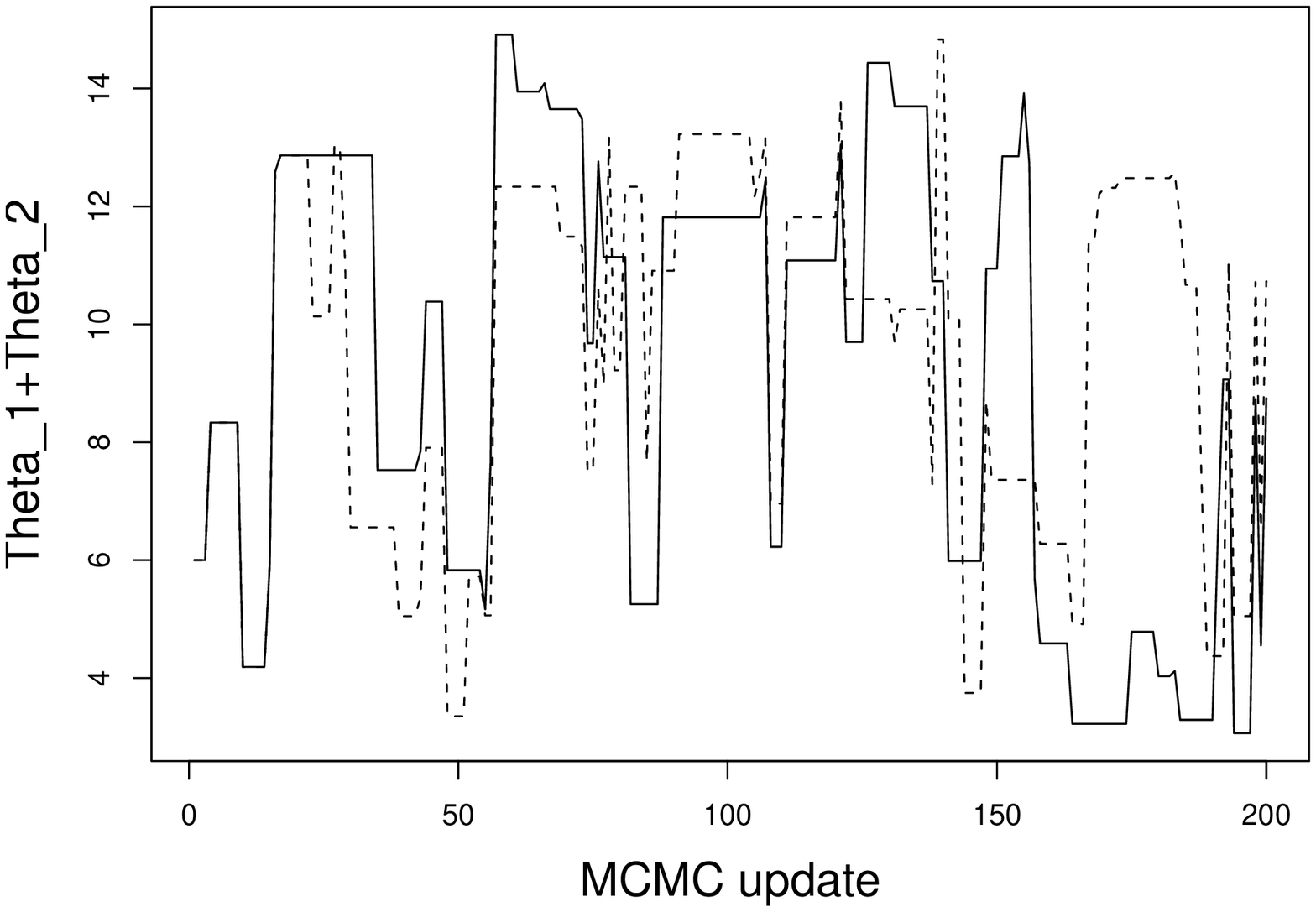}\\
   \includegraphics[width=4in]{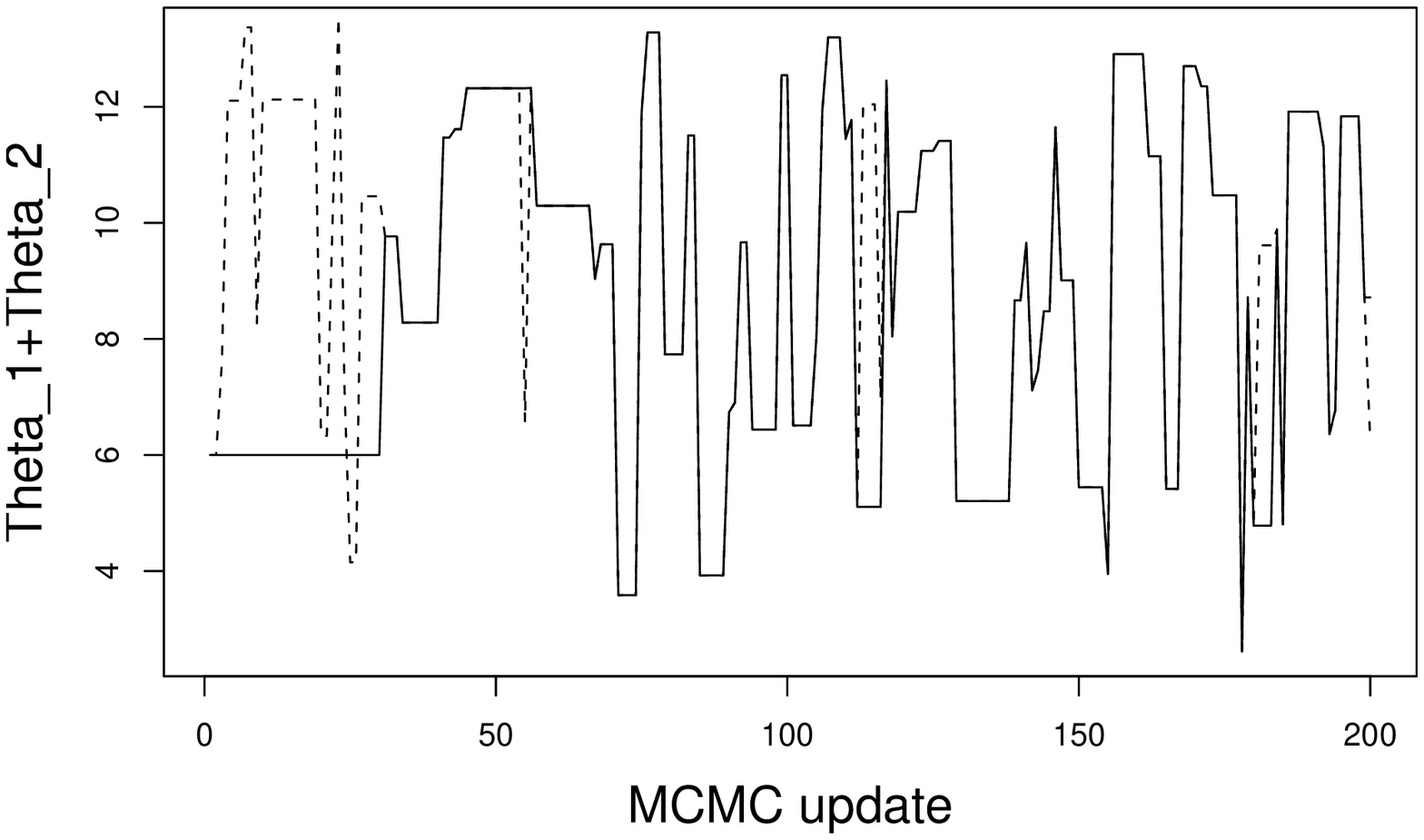}
  \caption{Simulations of the MCMC coupling-separation algorithm $\Theta_1+\Theta_2$ in the Penalty Method (solid lines) and Naive Algorithm
  (dashed lines), (Top) with Random-Walk Metropolis updates and (bottom) with Independence-Sampler updates. Target density
  is a mixture of bivariate normals, $D$-estimator using $m=8$ samples at each update.}\label{fig:coupling}
\end{figure}

In Fig~\ref{fig:coupling}
we plot realizations of coupled Penalty Method and naive MCMC samplers
using random walk Metropolis, and an Independence sampler.
The means of the mixture components are on $\theta_1=\theta_2$ so we plot $\Theta_1+\Theta_2$ against
MCMC update to show the mixing across components. For illustration, we compute our estimator $\hat D$ from
just eight independent $W_i$'s (we set $m=8$). This ensures that the mean
separation times are short, approximately 72 updates for the random walk update and 32 for the independence sampler.
This is convenient for graphical display.
In this simulation the random walk proposal (top) for the Penalty Method and naive samplers separate after 25 updates.
The walks shadow one another as they use the same uniform numbers to generate proposals.
The Independence samplers separate after just one update, but then coalesce, and go on to branch
and coalesce many times. In an Independence sampler run of length 10000 updates with $m=8$,
90\% of the samples returned by the naive sampler are exactly equal to those returned by the Penalty Method.

\begin{figure}
  % Requires
  \includegraphics[width=4in]{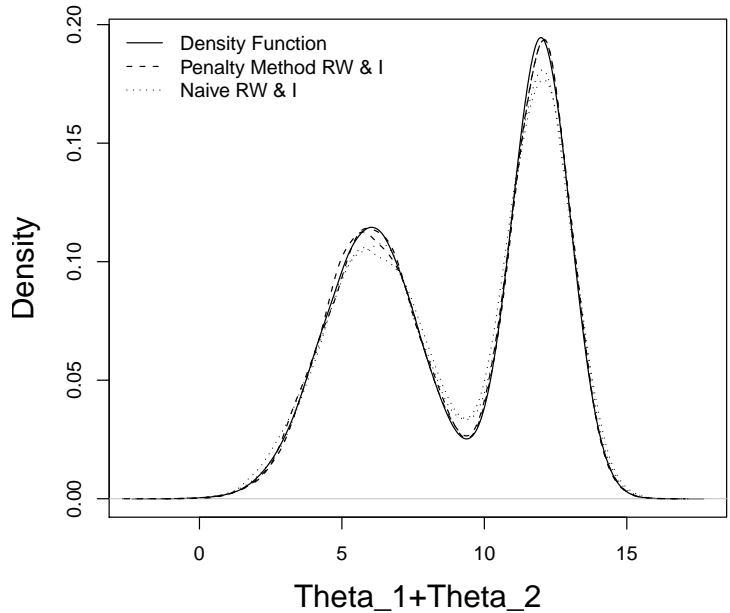}
  \caption{Density for $\Theta_1+\Theta_2$ in the target mixture of bivariate normals (solid curve) plotted with densities
  estimated using the Penalty Method (dashed curves for Independence and Random Walk updates) and Naive Algorithms (dotted curves).
  }\label{fig:density}
\end{figure}

In Fig~\ref{fig:density}
we show the slight over-dispersion
of the naive samplers compared to the target distribution.
This is visible because we used a very small sample size for the $D$-estimator $\hat D$ of $m=8$ iid samples.
In this example, the Penalty method
samplers are exact, even with $m=8$, and follow the target density to within error.

The upper graph
\begin{figure}
  \includegraphics[width=3.5in]{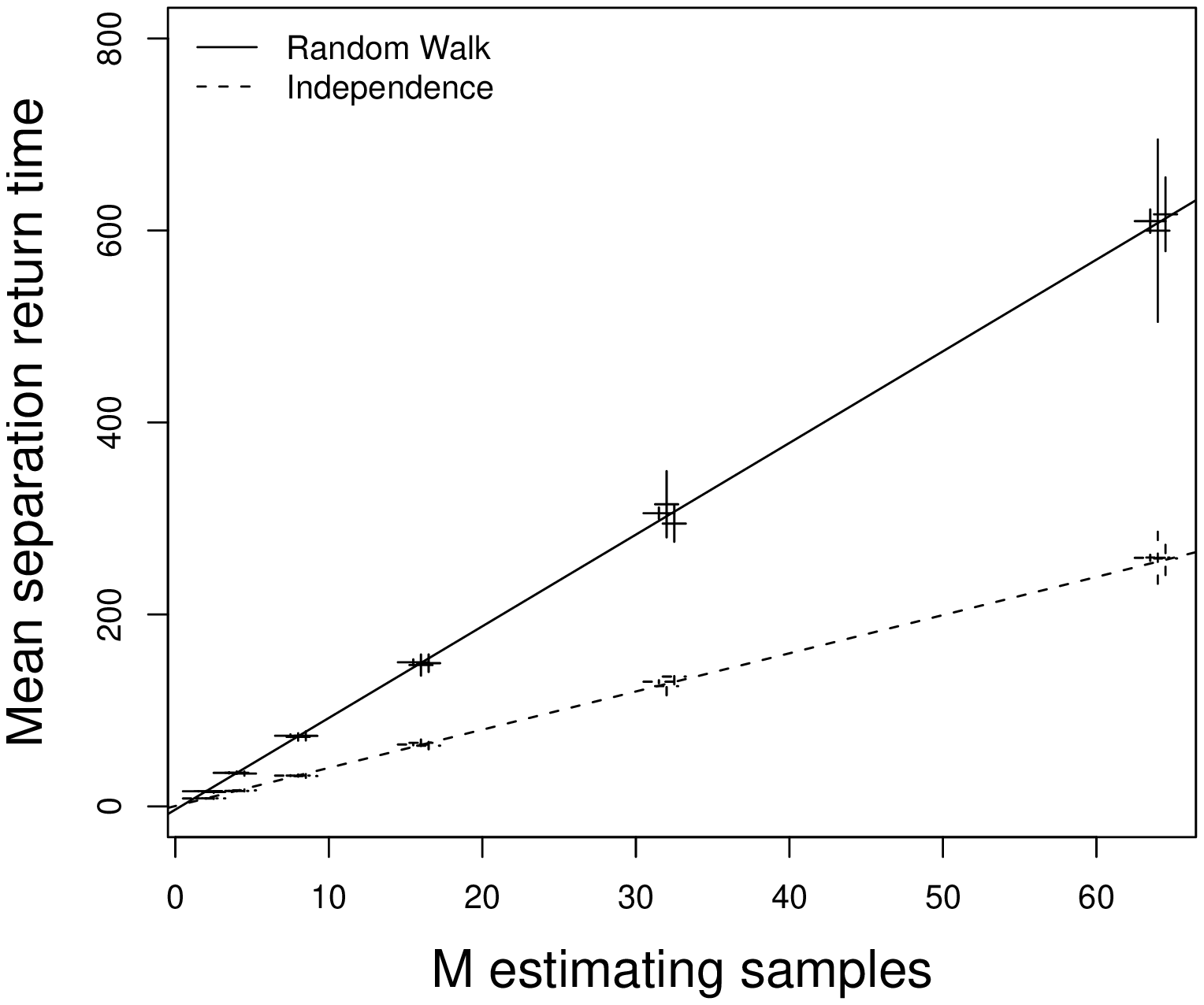}\\
  \includegraphics[width=3.5in]{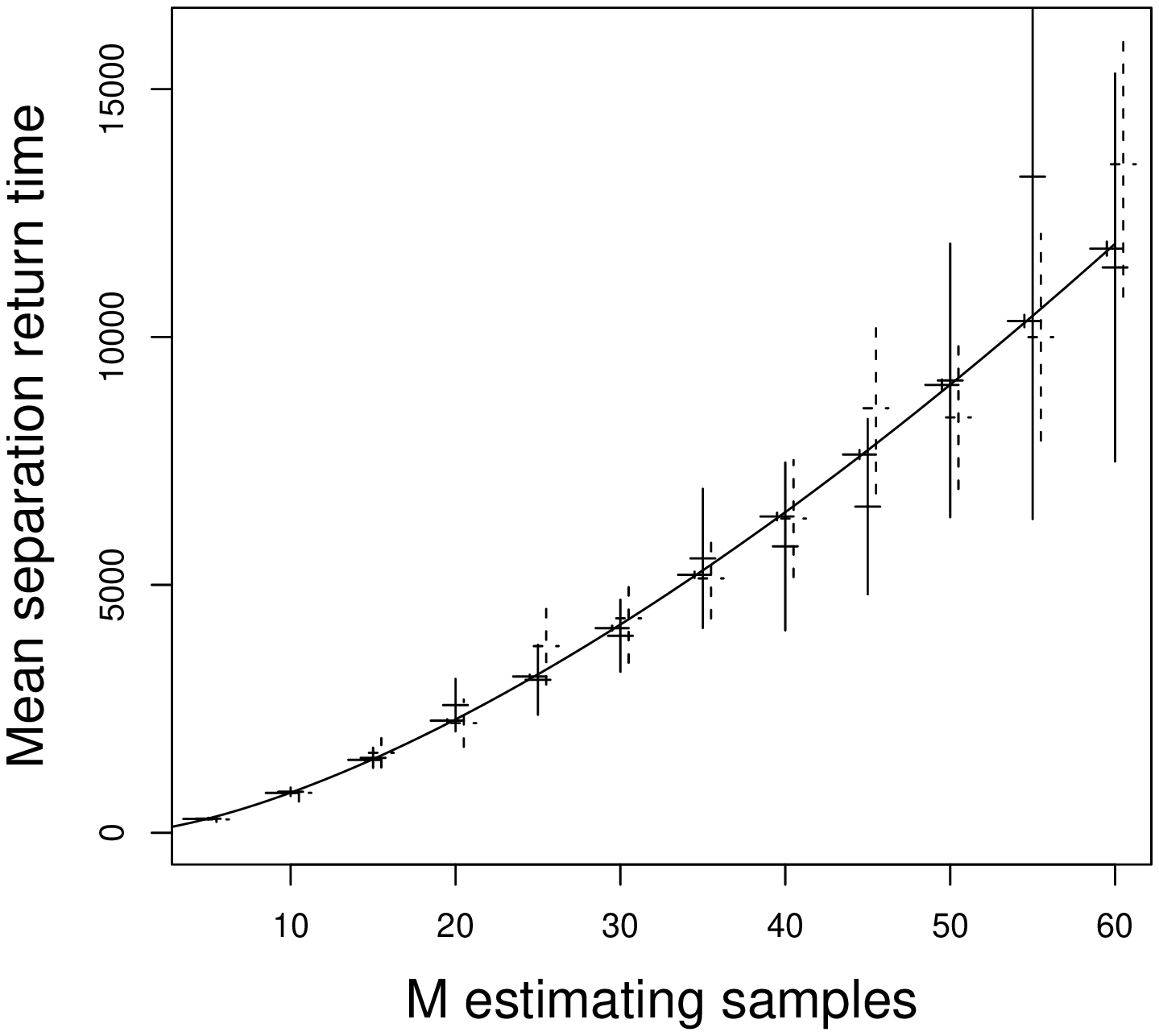}
  \caption{
  (Top) Estimated separation times $\rho$ and $\tau$ between the exact Penalty Method and the approximate Naive Algorithm, as a function of estimator sample size $m$, for Random-Walk (solid lines) and
  Independence sampler (dashed lines) updates.
  Two estimates of $\rho$, $\hat\rho_1$ (left error bar in each group of three) and $\hat\rho_2$ (central error bar) and $\hat\tau$
  (right error bar)
  are plotted for each sample and each $m$ with a linear regression of the
  $\hat\rho_2$ estimates. (Bottom) Estimated separation times, as above, between Penalty Method
  and approximate Penalty Estimate chains regressed with $\hat \rho_1=cm^{3/2}$.}\label{fig:septimes}
\end{figure}
in Fig~\ref{fig:septimes} demonstrates the linear dependence of the
separation return times $\rho$ and separation times $\tau$ on the number of samples $m$ used to form $\hat D$.
Two estimates of $\rho$ are computed. The first is from Eqn~\ref{eq:prt},
\[
\hat\rho_1=\frac{1}{K^{-1}\sum_{t=1}^K |\alpha_P(\theta_t,\theta'_t;y_t)-\alpha_N(\theta_t,\theta'_t;x_t)|}.
\]
For the second, $\hat\rho_2=S^{-1}\sum_{i=1}^S (T_i-T_{i-1})$ estimates $\rho$.
The $\hat\rho_1$ estimator has lower variance than the $\hat \rho_2$ estimator.
The $\tau$-estimator is the mean of 1000 realizations of $T$. The estimates are computed
for Random-Walk Metropolis and Independence Samplers. The Naive Independence sampler
separates more rapidly than the Naive Random-Walk sampler, on average, but can coalesce,
so it generates estimates of similar bias in this example. The separation return times $\rho$ and
separation times $\tau$ are approximately equal. The separation mark process $B_t$
is not in general Markov so $\mathbb{E}(T|B_0=1)\ne\mathbb{E}(T)$ in general. However in this example
the separation return times $T_i-T_{i-1}$ are not easily distinguished from iid geometric random numbers.

In order to demonstrate the $O(m^{3/2})$ behavior of separation times between coupled
Penalty Method and Penalty Estimate Method chains we take $W_i\sim N(0,1)$
$\hat D=D+\sum_i W_i/m$ an use $s^2(D_1,...,D_m)$ to estimate $\sigma^2$.
The estimate is unbiased and normal, but the variance is unknown. The separation time
function $\tau=\tau(m)$ displayed in the lower graph in Fig~\ref{fig:septimes} grows very rapidly with $m$.

\section{Discussion}

We have presented two new algorithms. The first is an `exact' MCMC algorithm for
a randomized acceptance probability, in which the randomization is not part of
the MCMC state. As this algorithm generalizes the simplest forms of
of the penalty method, universal algorithm and single variable exchange algorithms,
it may have applications in the simulation of doubly intractable target distributions.

Randomized MCMC is complementary to pseudo-marginal MCMC (Lin et al. (2000) \cite{lin00},
Beaumont (2003) \cite{beaumont03}, Andrieu et al (2009) \cite{andrieu09}) and Andrieu et al (2010) \cite{andrieu10}.
In both algorithms an intractable target distribution is estimated using auxiliary random variables.
However, in randomized MCMC algorithms the auxiliary variables extend the transition kernel, and it is
the transition kernel which has the correct marginal at each update (ie, integrating Eqn~\ref{eq:vdb}),
whereas in pseudo-marginal algorithms, the auxiliary variables extend the target distribution,
and the target distribution has the correct marginal distribution. Randomized MCMC does not
maintain auxiliary variables in the state, and this may be an advantage. Beaumont (2003)\cite{beaumont03}
notes that pseudo-marginal algorithms are prone to getting stuck when the target density is over-estimated.
Simulation
studies using very simple target densities and comparable exact pseudo-marginal and penalty method
algorithm show the efficiency advantage of the Penalty Method increasing with increasing estimator variance.
Against this, existing exact randomized algorithms impose much stronger conditions on the distribution of
the estimator than pseudo-marginal algorithms.

In fact, the class of randomized Metropolis Hastings MCMC algorithms we have described
are themselves special cases of a very large class of MCMC algorithms called
Active Particle Algorithms, described in Lee et al. (2011) \cite{lee11}.
This class of algorithms includes the pseudo-marginal MCMC and randomized algorithms
as special cases, as well as mixed algorithms in which both the target distribution and
transition kernel in detailed balance are both marginals.

The second algorithm, the coupling-separation algorithm, is not
directly useful for doubly intractable problems, as the simple form of the exact penalty method
algorithm is not tractable. However, it shows that under some conditions the naive algorithm generates
the exact same sequence of samples as
an exact penalty method algorithm, out to $O(m)$ steps in the MCMC, where $m$
is the sample size used to estimate the log of the Hastings ratio.
This suggests strategies for improving naive simulation, using any method
that tends to increase this separation time. The coupling-separation algorithm may help
to suggest improvements in other settings involving an approximate likelihood in
an MCMC algorithm. As an example, if we have a very large data set, and a
log-likelihood given as sum over independent data, we may estimate the log-likelihood
using a small sample of size $m$ drawn with replacement from the data. The estimator is asymptotically normal in $m$.
%This will work if the data are fairly homogeneous.

The coupling-separation algorithm complements the perfect simulation
algorithm of Propp and Wilson. In that algorithm coupled MCMC chains for the same target
start in different states and coalesce. In the coupling-separation algorithm coupled MCMC chains
for different targets start in the same state and branch.

\subsection*{Appendix A: Peskun ordering of s-chains and r-chains}

We now prove the ordering given in Eqn~\ref{eq:tplb}.
Let
\[
A=\{x\in W: \xi(x;\theta,\theta')<h(\theta,\theta')\xi(f(x);\theta',\theta)|f'(x)|\},
\]
so that %$\alpha_\xi(\theta,\theta';x)=1$ for $x\in A$.
\begin{equation}\label{eq:apip}
  \alpha_\xi(\theta,\theta') = \int_A \xi(x;\theta,\theta') dx + \int_{W\setminus A} \!\! h(\theta,\theta')\xi(f(x);\theta',\theta)|f'(x)| dx.
\end{equation}
Replacing $\xi(x;\theta,\theta')$ with $h(\theta,\theta')\xi(f(x);\theta',\theta)|f'(x)|$ for $x$ in $A$,
\begin{eqnarray*}
  \alpha_\xi(\theta,\theta') &\le& \int_{W} h(\theta,\theta')\xi(f(x);\theta',\theta)|f'(x)| dx \\
    & = & h(\theta,\theta'),
\end{eqnarray*}
while replacing $h(\theta,\theta')\xi(f(x);\theta',\theta)|f'(x)|$ with $\xi(x;\theta,\theta')$ for $x$ in $W\setminus A$,
\begin{eqnarray*}
  \alpha_\xi(\theta,\theta') &\le& \int_{W} \xi(x;\theta,\theta') dx \\
    & = & 1.
\end{eqnarray*}
It follows that $\alpha_\xi(\theta,\theta')\le \min(1, h(\theta,\theta'))$.

\subsection*{Appendix B: irreducibility and minorization of r-chains}

We now prove that Eqn~\ref{eq:tplb} follows from Eqn~\ref{eq:epsbound}.
Let
\[
\Xi(X)=\frac{\xi(f(X);\theta',\theta)}{\xi(X;\theta,\theta')}|f'(X)|
\]
so that $\alpha_\xi(\theta,\theta';X)=\min(1,h(\theta,\theta')\Xi(X))$ and $\Xi$ depends on $\theta$ and $\theta'$.
From the definition of $\alpha_\xi(\theta,\theta')$ in Eqn~\ref{eq:apxi},
\begin{eqnarray}\label{eq:lbxi}
\alpha_\xi(\theta,\theta') &=& \mathbb{E}\left(\alpha_\xi(\theta,\theta';X)\right)\\\nonumber
                            &=& \mathbb{E}\left(\alpha_\xi(\theta,\theta';X)\vert \Xi(X)\ge 1\right)\Pr(\Xi(X)\ge 1)\quad +\\\nonumber
                            && \mathbb{E}\left(\alpha_\xi(\theta,\theta';X)\vert \Xi(X)< 1\right)\Pr(\Xi(X)< 1)\\\nonumber
    &\ge& \mathbb{E}\left(\alpha(\theta,\theta')\vert \Xi(X)\ge 1\right)\Pr(\Xi(X)\ge 1)\\\nonumber
    &=& \alpha(\theta,\theta')\Pr(\Xi(X)\ge 1)
\end{eqnarray}
where the inequality in the third equation holds because $\alpha_\xi(\theta,\theta';X)\ge \alpha(\theta,\theta')$ given $\Xi(X)\ge 1$.

Let
\begin{equation}\label{eq:tpxi}
p_\xi(\theta,\theta')=\left\{\begin{array}{cc}
                           q(\theta,\theta')\alpha_\xi(\theta,\theta') & \theta\ne \theta' \\
                           0 & \theta=\theta'
                         \end{array}\right.
\end{equation}
be the zeroed transition probability density
in the r-algorithm.
The condition in Eqn~\ref{eq:epsbound} gives $\Pr(\Xi(X)\ge 1)\ge \epsilon$ for some $\epsilon>0$ not depending on $\theta,\theta'$,
and so $p_\xi(\theta,\theta')\ge \epsilon p(\theta,\theta')$ for the transition density, from Eqns~\ref{eq:tp}, \ref{eq:lbxi} and \ref{eq:tpxi}.
Let
\[
r_\xi(\theta)=1-\int_E p_\xi(\theta,\theta')\mu(d\theta')
\]
give the probability to remain in the same state, in the r-algorithm.
By the Peskun ordering, $r_\xi(\theta)\ge r(\theta)$, and so trivially, $r_\xi(\theta)\ge \epsilon r(\theta)$, and hence
\begin{eqnarray*}
P_\xi(\theta,B) &=& \int_B p_\xi(\theta,\theta')\mu(d\theta')+r_\xi(\theta)\mathbb{I}_{\theta\in B} \\
   &\ge& \epsilon P(\theta,B)
\end{eqnarray*}
for each $\theta\in E$ and $B\in \mathcal E$. This is Eqn~\ref{eq:tplb} and we are done.

If $\xi$ does not depend on $\theta,\theta'$ then condition Eqn~\ref{eq:epsbound} is not needed. In that case $0<\Pr(\Xi(X)\ge 1)<1$, since if $$\xi(f(x))|f'(x)|<\xi(x)$$
for $\xi$-a.e. $x\in W$, then the two functions cannot both be probability densities, which is a contradiction.
This implies Eqn~\ref{eq:epsbound} as there is no $\theta,\theta'$-dependence in $\Pr(\Xi(X)\ge 1)$.
If there is $\theta'$ dependence, as in the single variable exchange algorithm in Section~\ref{sec:examples},
then we need to exclude cases where $\Pr(\Xi(X)\ge 1)\rightarrow 0$ as $\theta'$ approaches a boundary of $E$,
in order that Eqn~\ref{eq:tplb} hold for $\epsilon$ independent of $\theta$ and $C$.

\bibliographystyle{plain}
\bibliography{NichollsFoxMuirWatt12v1}

\begin{thebibliography}{10}

\bibitem{andrieu10}
C.~Andrieu, A.~Doucet, and R.~Holenstein.
\newblock Particle {M}arkov chain {M}onte {C}arlo methods.
\newblock {\em J. Roy. Statist. Soc Series B}, 72:269--342, 2010.

\bibitem{andrieu09}
Christophe Andrieu and Gareth~O. Roberts.
\newblock The pseudo-marginal approach for efficient {M}onte {C}arlo
  computations.
\newblock {\em Ann. Statist.}, 27(2):697--725, 2009.

\bibitem{Ball2003}
Robin Ball, Thomas Fink, and Neill Bowler.
\newblock Stochastic annealing.
\newblock {\em Physical Review Letters}, 91(3):1--4, July 2003.

\bibitem{beaumont03}
M.~A. Beaumont.
\newblock Estimation of population growth or decline in genetically monitored
  populations.
\newblock {\em Genetics}, 164:1139--1160, 2003.

\bibitem{besag95}
J.~E. Besag, P.~J. Green, D.~Higdon, and K.~Mengersen.
\newblock {B}ayesian computation and stochastic systems.
\newblock {\em Statist. Sci.}, 10:3--66, 1995.

\bibitem{Ceperley1999}
D~M Ceperley and M~Dewing.
\newblock {The Penalty Method for Random Walks with Uncertain Energies}.
\newblock {\em Journal of Chemical Physics}, 110, 1999.

\bibitem{gordon93}
N.~J. Gordon, D.~J. Salmond, and A.~F.~M. Smith.
\newblock Novel approach to nonlinear/non-{G}aussian {B}ayesian state
  estimation.
\newblock {\em IEE Proceedings-F on Radar and Signal Processing},
  140(2):107--113, 1993.

\bibitem{hastings70}
W.~Hastings.
\newblock {M}onte {C}arlo sampling methods using {M}arkov chains and their
  applications.
\newblock {\em Biometrika}, 57:97--109, 1970.

\bibitem{Kontoyiannis03}
I.~Kontoyiannis and S.~P. Meyn.
\newblock Spectral theory and limit theorems for geometrically ergodic {M}arkov
  processes.
\newblock {\em The Annals of Applied Probability}, 13:304–362, 2003.

\bibitem{lee11}
A.~Lee, C.~Andrieu, and A.~Doucet.
\newblock An active particle perspective of {MCMC} and its application to
  locally adaptive {MCMC} algorithms.
\newblock 2011.

\bibitem{lin00}
L.~Lin, K.~F. Liu, and J.~Sloan.
\newblock A noisy {M}onte {C}arlo algorithm.
\newblock {\em Phys. Rev. D}, 61:074505, Mar 2000.

\bibitem{metropolis53}
N.~Metropolis, A.~W. Rosenbluth, M.~N. Rosenbluth, A.~H. Teller, and E.~Teller.
\newblock Equation of state calculations by fast computing machines.
\newblock {\em Journal of {C}hemical {P}hysics}, 21:1087--1092, 1953.

\bibitem{Moller2004}
Jesper M{\o}ller, A.N. Pettitt, K.~K. Berthelsen, and R.~W. Reeves.
\newblock {An efficient {M}arkov chain {M}onte {C}arlo method for distributions
  with intractable normalising constants}.
\newblock {\em Biometrika}, 93(2), June 2004.

\bibitem{murdoch98}
D.~J. Murdoch and P.~J. Green.
\newblock Exact sampling from a continuous state space.
\newblock {\em Scandinavian Journal of Statistics}, 25:483--502, 1998.

\bibitem{Murray2006}
Iain Murray and David J~C Mac{K}ay.
\newblock {MCMC} for doubly-intractable distributions.
\newblock In {\em Proceedings of the 22nd Annual Conference on Uncertainty in
  Artificial Intelligence (UAI)}, 2006.

\bibitem{peskun73}
P.H. Peskun.
\newblock Optimum {M}onte {C}arlo sampling using {M}arkov chains.
\newblock {\em Biometrika}, 60:607--612, 1973.

\bibitem{propp98}
James~G. Propp and David~B. Wilson.
\newblock How to get a perfectly random sample from a generic {M}arkov chain
  and generate a random spanning tree of a directed graph.
\newblock {\em Journal of Algorithms}, 27:170--217, 1998.

\bibitem{propp96}
J.G. Propp and D.B. Wilson.
\newblock Exact sampling with coupled {M}arkov chains and applications to
  statistical mechanics.
\newblock {\em Random Structures and Algorithms}, 9:223--252, 1996.

\bibitem{tierney94}
L.~Tierney.
\newblock Markov chains for exploring posterior distributions.
\newblock {\em Annals of Statistics}, 22:1701--1762, 1994.

\end{thebibliography}
\end{document}